\documentclass[journal]{IEEEtran}
\usepackage{mathrsfs}
\usepackage{url,times,amsmath,color,amssymb,graphicx,epsfig,psfig,cite,psfrag,subfigure,dsfont,booktabs}
\usepackage{algorithm}
\usepackage{graphicx}
\usepackage{algorithmic}
\usepackage{bm}
\usepackage{color}
\usepackage{multicol}
\usepackage{stfloats}

\begin{document}
\title{Energy Efficient User Clustering, Hybrid Precoding and Power Optimization in Terahertz MIMO-NOMA Systems}

\author{Haijun Zhang,~\IEEEmembership{Senior Member,~IEEE}, Haisen Zhang, Wei liu, Keping Long,\\~\IEEEmembership{Senior Member,~IEEE}, Jiangbo Dong, Victor C. M. Leung,~\IEEEmembership{Fellow,~IEEE}

\thanks{This work is supported by the National Natural Science Foundation of China (61822104, 61771044), Beijing Natural Science Foundation (L172025, L172049), 111 Project (No. B170003), and the Fundamental Research Funds for the Central Universities(FRF-TP-19-002C1, RC1631), Beijing Top Discipline for Artificial Intelligent Science and Engineering, University of Science and Technology Beijing.  This paper was presented in part at the IEEE International Conference on Communications (ICC 2020), Dublin, Ireland, 2020. The corresponding authors are Keping Long and Haijun Zhang.

Haijun Zhang, Haisen Zhang, and Keping Long are with Institute of Artificial Intelligence, Beijing Advanced Innovation Center for Materials Genome Engineering, Beijing Engineering and Technology Research Center for Convergence Networks and Ubiquitous Services, University of Science and Technology Beijing, Beijing 100083, China (e-mail: haijunzhang@ieee.org, z-haisen@qq.com, longkeping@ustb.edu.cn).

Wei liu and Jiangbo Dong are with China Mobile Group Design Institute Co., Ltd, Beijing, 100080, China.

Victor C. M. Leung is with the College of Computer Science and Software Engineering, Shenzhen University, Shenzhen 518060, China, and also with the Department of Electrical and Computer Engineering, the University of British Columbia, Vancouver, BC V6T 1Z4, Canada (e-mail: vleung@ieee.org).

}} \maketitle

\thispagestyle{empty} 
\pagestyle{empty}

\begin{abstract}
Terahertz (THz) band communication has been widely studied to meet the future demand for ultra-high capacity.
In addition, multi-input multi-output (MIMO) technique and non-orthogonal multiple access (NOMA)  technique with multi-antenna also enable the network to carry more users and provide multiplexing gain.
In this paper, we study the maximization of energy efficiency (EE) problem in THz-NOMA-MIMO systems for the first time.
And the original optimization problem is divided into user clustering, hybrid precoding and power optimization.
Based on channel correlation characteristics, a fast convergence scheme for user clustering in THz-NOMA-MIMO system  using enhanced K-means machine learning algorithm is proposed.
Considering the power consumption and implementation complexity, the hybrid precoding scheme based on the sub-connection structure is adopted.
Considering the fronthaul link capacity constraint, we design a distributed alternating direction method of multipliers (ADMM) algorithm for power allocation to maximize the EE of THz-NOMA cache-enabled system with imperfect successive interference cancellation (SIC).
The simulation results show that the proposed user clustering scheme can achieve faster convergence and higher EE, the design of the hybrid precoding of the sub-connection structure can achieve lower power consumption and power optimization can achieve a higher EE for the THz cache-enabled network.
\end{abstract}

\section{Introduction}

The terahertz (THz) band is located between infrared and microwave, and the frequency range is 0.1 THz-10 THz.
Using THz wave as carrier signal of wireless communication, high-speed broadband wireless communication is one of the most interesting problems of THz \cite{JSACTHz}.
The THz band is 3-4 orders of magnitude higher than the current wireless communication band commonly used in mobile phones, which can provide huge communication bandwidth \cite{highspeedTHz}.
THz communication has the advantages of large capacity, good direction, strong confidentiality and strong anti-interference ability.
So the THz wave based high data rate short-range broadband wireless communication is feasible.
In the terrestrial  wireless communication, THz communication can obtain tens of Gb/s wireless transmission rate, which is significantly better than the current ultra-wideband technology \cite{THzintro}.
Compared with low-frequency wireless channel, although THz channel has larger free transmission attenuation loss and atmospheric molecule and water droplet absorption attenuation, high-speed communication in THz band can still be achieved in short distance by improving the gain of transmitting antenna and receiving antenna.
In order to achieve the communication rate of 10 Gb/s or higher in the future, non-orthogonal multiple access (NOMA) and multi-input multi-output (MIMO) technology \cite{MIMO,z1,z3} in the existing microwave communication technology are also needed to improve the energy efficiency (EE).

Compared with the orthogonal multiple access (OMA) system, NOMA enables wireless communication to serve more user loads and requirements, which can improve the system throughput.
In addition, weak users with successive interference cancellation (SIC) can weaken the interference from strong users \cite{B5GNOMA}.
In particular, the combination of NOMA and THz band enables a large number of antennas to be integrated into the chips with acceptable size due to the small wavelength of THz.
Furthermore, the severely strong correlation channel promotes multiple users to work in the same beam, thus greatly reducing the hardware requirements and power consumption of RF, which is also the foundation for user clustering.

Furthermore, the precoding of MIMO system has been widely studied in low frequency band as a key technology to improve wireless communication capacity \cite{HP2019}.
In particular, precoding in THz network requires lower complexity and larger antenna size.
In practice, pure digital precoding is often not feasible, because it requires higher baseband processing capacity, so hybrid precoding applied to THz-NOMA system needs to be studied.
Several precoding schemes in THz have been proposed in \cite{THzOne-Bit,THz2019P,JSAC1bit}.
The authors of \cite{THzOne-Bit} studied the hybrid precoding scheme in THz system, including the maximum ratio transmission (MRT) and zero forcing (ZF) precoding.
In \cite{THz2019P}, the authors studied a single carrier precoding and detection algorithm for frequency selective THz channels.
Moreover, the capacity characteristics and precoding of multiple-input single-output (MISO) fading channels of one-bit transceiver are studied in \cite{JSAC1bit}.
However, these studies are far from satisfying the practical application of THz-NOMA network.

The large band of THz network will serve huge amount of data services, which will not only cause heavy transmission burden, but also leads to energy cost. Caching in the BS side is an effective solution to reduce transmission burden\cite{d4}.
And many precoding schemes for cached-enabled system have been proposed.
The authors of \cite{c2017} believed that cache played an important role in low power efficient backhaul link networks.
The precoding problem for MIMO wireless access network supporting caching was studied in \cite{cUD2017}.
However, caching in THz-NOMA networks still needs further study. Because the communication capacity of THz network is much larger than that of the existing wireless access network, which makes the system more complex, such as the capacity limitation of the fronthaul link.

Furthermore, although the THz band based NOMA caching system can significantly improve data rate, it is more difficult to design user clustering, precoding, and power optimization.
In addition, the heavy attenuation of THz band  and the small transmit power of THz transmitter make THz wave different with the other frequency band based wireless networks.
Therefore, it is necessary to study the power optimization scheme for THz networks to solve the problems of low transmit power.
In \cite{THzbpb2019}, a downlink THz-NOMA system was proposed, and the beamforming and resource optimization problems were studied in order to guarantee the quality of service (QoS) requirement of users. However, this study does not consider the effect of channel correlation for users and the EE of the system is not measured.
There are a lot of studies focusing on user clustering and power optimization in low-frequency band networks \cite{Noma2}\cite{DLL2019}.
Although there are a lot of studies focusing on user clustering and power optimization in low-frequency band networks, whether these schemes are suitable for THz band networks is still need to be studied.
Therefore, the system model of resource management in THz-NOMA networks is challenging and difficult to be solved.

To the best of our knowledge, the resource optimization has not been well investigated  in THz-NOMA networks.
In this paper, THz band and NOMA technologies are used for the communications between small cell base station (SBS) and users.
We focus on the resource optimization problem of user clustering, hybrid precoding, and power optimization to maximize the system energy efficiency  in a downlink heterogeneous THz-NOMA-MIMO network.
Reference \cite{ICC2020} is a conference version of this paper. We extend \cite{ICC2020} in the following ways: (1) the power allocation algorithm is presented now; (2) we provide complexity analysis for the proposed algorithms; (3) more simulation results are provided to verify the proposed methods.
The main contributions of this paper can be summarized as follow:
\begin{itemize}
  \item Firstly, a physical channel model for THz-NOMA heterogeneous downlink cache-enabled system is provided.
      Due to the weak scattering ability of THz band communication and its sensitivity to channel congestion, we consider a line-of-sight (LOS) link whose path loss include spreading loss and molecular absorption loss.
      In addition, the system EE performance model combined with caching model and power consumption model is given.
  \item In this paper, we study the problem of EE maximization  in THz-NOMA-MIMO systems and design user clustering, hybrid precoding and  power optimization schemes. We divide the original optimization problem into three sub-problems and solving them separately.
  \item A fast convergence scheme for user clustering in NOMA-MIMO system by using enhanced K-means machine learning algorithm is proposed, which is based on channel correlation characteristics.
      Considering the power consumption and implementation complexity, we adopt a hybrid precoding scheme based on the sub-connection structure of the quantized phase shifter and the low complexity ZF algorithm.
  \item  Due to the limited transmitting power and error propagation of practical THz system, we derive a new expression of data rate for residual interference in imperfect SIC case. Considering the fronthaul link capacity constraint, we design a distributed alternating direction method of multipliers (ADMM) algorithm for power allocation to maximize the EE of THz system.
  \item The simulation results show that the proposed user clustering scheme can achieve faster convergence and higher EE for THz-NOMA-MIMO system.
      At the same time, the design of the hybrid precoding of the sub-connection structure can achieve lower power consumption, and power optimization can achieve a higher EE for the THz cache-enabled network.
\end{itemize}

The rest of the paper is organized as follows.
Section II shows the system model of THz-NOMA network and formulate the system EE optimization problem.
In section III, user clustering and hybrid precoding are designed carefully.
In section IV, power optimization for THz network is proposed.
The extensive simulation results analysis are discussed to illustrate THz-NOMA-MIMO system EE performance in section V. Finally, we summarize the paper in section VI.

\emph{Notation:}  Bold capital letters represent matrices, and bold lower case letters represent vectors. We use ${\left[  \cdot  \right]^H}$ and  ${\left\|  \cdot  \right\|_p}$ to represent conjugate transposition and $l_p$ norm of matrices, respectively. And $j$ is the imaginary unit, i.e., $j^2 =-1$.

\section{System Model And Problem Formulation}

\begin{figure}[t]
        \centering
        \includegraphics*[width=8cm]{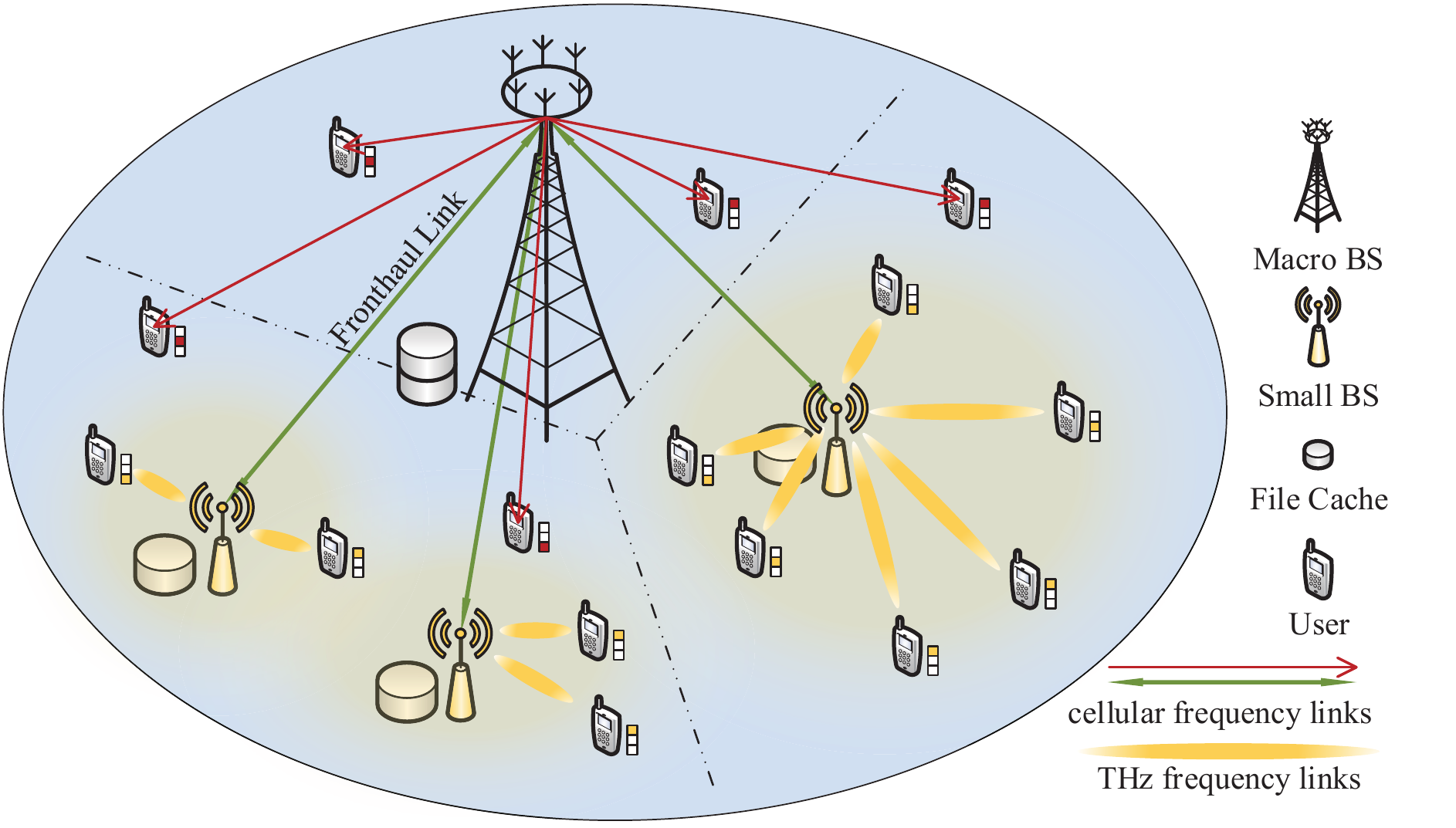}
       \caption{Cache-Enabled THz-NOMA-MIMO  Downlink Network Architecture.}
        \label{fig:111}
\end{figure}

In this paper, we consider a downlink heterogeneous THz NOMA-MIMO network, where the macro base station (MBS) and SBSs are equipped with $N_T$ antennas and $N_R$ RF and transmit superimposed signal information through NOMA technology on each beam. The set of all BS is represented as $\mathcal{B} = \left\{ {1,2, \cdots ,b, \cdots ,B,B + 1} \right\}$, where $B+1$ is the MBS and the rest are SBS.
As shown in Fig. \ref{fig:111}, each BS is equipped with a cache storage and connected to the central controller via an error-free fronthaul link of capacity $\bm{C}^{FH}_b$ bit/symbol. And the set of all user is represented as $U = \left\{ {{U_1},{U_2}, \cdots ,{U_b}, \cdots ,{U_{B + 1}}} \right\}$.
All users are randomly distributed within the coverage of the MBS cell.
Unlike the MISO systems, all single-antennas users are served in the form of clusters.
${U_b} = \left\{ {u_{b,1}^1,u_{b,1}^2, \cdots ,u_{b,N}^1,u_{b,N}^2,u_{b,N}^ {\cdots} } \right\}$
 indicates that each user is divided into $N$ clusters, where $u_{b,n}^i$ is the $i$th user in $n$th cluster of $b$th BS.
In this study, the orthogonal time-frequency resources are allocated to the clusters served by each BS. We assume that all SBS have perfect channel state information (CSI) for all users, and superimpose transmission signals on each beam of each BS through NOMA technology.

Consider the MIMO-NOMA communication system employing a two-tier heterogeneous network, MBS serves users $U_{B + 1}$, and all SBSs receive the data of users $ \left\{ {{U_1},{U_2}, \cdots ,{U_B}} \right\}$ (i.e. fronthaul data) from MBS.
MBS uses cellular network frequency band because of the need for wide coverage. However, SBSs usually have a smaller coverage radius, so THz band can be used to provide short-distance and high-speed services.
Then the cross-tier interference can be ignored due to different bands.
Moreover, due to the huge data rate of THz band network, the capacity of cellular frequency BS is much smaller than that of THz band BS, this paper will focus on the utility of THz band BS.

\subsection{Hybrid Analog/Digital Precoding Model}

\begin{figure}[t]
        \centering
        \includegraphics*[width=8cm]{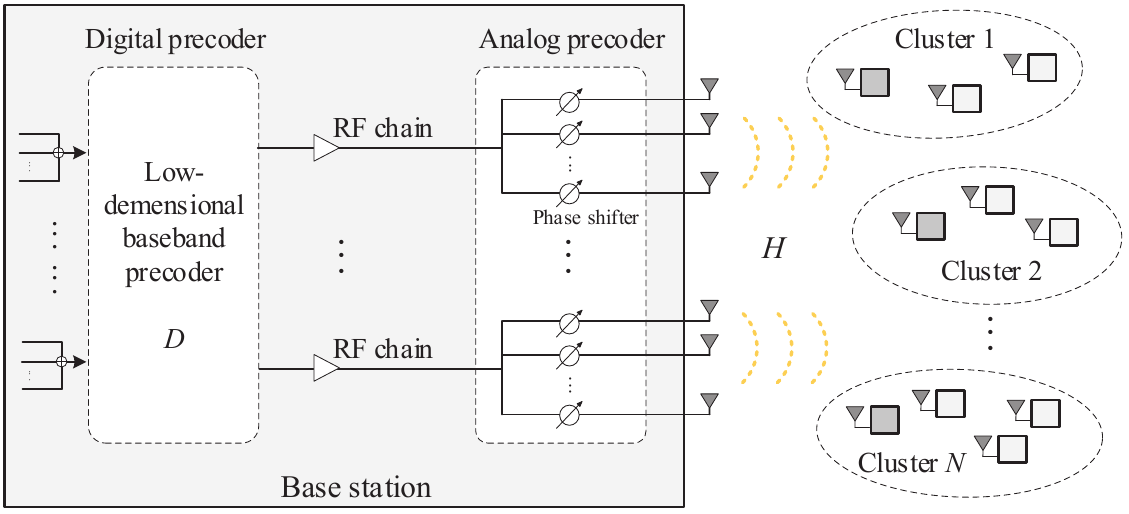}
       \caption{System model of sub-connected HP architecture for NOMA-MIMO Network.}
        \label{fig:222}
\end{figure}

To reduce energy consumption, we consider the sub-connected structure, where each RF chain is connected to only part of the antennas via phase shifter and the number of phase shifters is equal to the number of antennas as seen in Fig. \ref{fig:222}.
Herein, the number of antennas connected to each RF chain is equal, expressed as $N_T^{sub} = N_T / N_R$ and $N_T^{sub}$ should be an integer.
In order to achieve multiplexing gain, we make the number of RF chains equal to the number of clusters, which has been adopted by many study \cite{THzOne-Bit}\cite{DLL2019}.

Based on the system model and parameters mentioned above, the signals of the users in $n$th cluster of BS $b$ are superimposed as follows:
\begin{equation}
{\bm{x}_{b,n}} = \sum\limits_{i = 1}^{{L_{b,n}}} {\sqrt {\rho _{b,n}^i{p_{b,n}}} \bm{s}_{b,n}^i}
\end{equation}
where ${\rho _{b,n}} = \left\{ {\rho _{b,n}^1,\rho _{b,n}^2, \cdots ,\rho _{b,n}^{{L_{b,n}}}} \right\}$ denotes the power control factor for users of $n$th cluster in BS $b$, ${\bm{s} _{b,n}} = \left\{ {\bm{s} _{b,n}^1,\bm{s} _{b,n}^2, \cdots ,\bm{s}_{b,n}^{{L_{b,n}}}} \right\}$ is the signal set of users,  $p_{b,n}$ is the transmit power of $n$th cluster in BS $b$, $L_{b,n}$ is the number of users of $n$th cluster in BS $b$. The  transmit power for each cluster is limited as
\begin{equation}
\sum\limits_{n = 1}^N {{{p_{b,n}}}  \le {p_b},\forall } b \in \mathcal{B}
\end{equation}
And the power control factor at $n$th cluster of BS $b$ is limited as \begin{equation}
\sum\limits_{i = 1}^{{L_{b,n}}} { {\rho _{b,n}^i}  = 1,\forall } b \in \mathcal{B},n \in N
\end{equation}

Next, the superposed signal is precoded by baseband precoder and phase shifters. Then, the user receives the signal transmitted by the antenna through the THz channel $\bm{H}$. $\bm{H} = \{ \left. {{\bm{H}_1},{\bm{H}_2}, \cdots ,{\bm{H}_b}, \cdots ,{\bm{H}_{B }}} \right|{\bm{H}_b} = \{ \bm{h}_{b,n}^1, \cdots ,\bm{h}_{b,N}^{{L_{b,N}}}\} \} $ is the $B\times{N}\times{U_b}$ channel matrix.
Thus, the received signal of $i$th user at $n$th cluster in BS $b$ can be given by
\begin{equation}
\bm{y}_{b,n}^i = \bm{h}_{b,n}^i{\bm{A}_b}\sum\limits_{n = 1}^N {{\bm{d}_{b,n}}} {\bm{x}_{b,n}} + \bm{\upsilon}_{b,n}^i
\end{equation}
where $\bm{A}_b$ is the analog precoding matrix, $\bm{d}_{b,n}$ is the digital precoding vector, $\bm{\upsilon} _{b,n}^i \sim CN(0,\sigma _{b,n}^2)$ is the AWGN with zero mean and variance $\sigma _{b,n}^2$.
For sub-connected structure, $\bm{A}_b$ forms as 
\begin{equation}
{\bm{A}_b} = \left[ {\begin{array}{*{20}{c}}
{\bm{a}_{b,1}^{sub}}&0& \cdots &0\\
0&{\bm{a}_{b,2}^{sub}}& \cdots &0\\
 \vdots & \vdots & \ddots & \vdots \\
0&0& \cdots &{\bm{a}_{b,{N_R}}^{sub}}
\end{array}} \right]
\end{equation}
where each ${\bm{a}_{b,n}^{sub}}$ at $n$ RF chain shares the same amplitude $\frac{1}{{\sqrt {N_T^{sub}} }}$.

Due to the huge loss of THz channel characteristics, it is difficult for users to receive signal from other SBSs, so the impact between SBS is very tiny, and the intra-tier interference between SBS can be neglected.
In the THz-NOMA downlink system, the interference in THz system includes intra cluster interference (ICI) and multi cluster interference (MCI).
Further, the received signal of $i$th user at $n$th cluster in BS $b$ is represented as
\begin{equation}
\begin{aligned}
& \bm{y}_{b,n}^i =
& & { \underbrace {\bm{h}_{b,n}^i{\bm{A}_b}{\bm{d}_{b,n}}\sqrt {\rho _{b,n}^i{p_{b,n}}} \bm{s}_{b,n}^i}_{{\rm{desired \ signal}}} +  \underbrace {\bm{h}_{b,n}^i{\bm{A}_b}\sum\limits_{j = 1,j \ne n}^N {{\bm{d}_{b,j}}} {\bm{x}_{b,j}}}_{{\rm{MCI}}} } \\
&&& { + \underbrace {\bm{h}_{b,n}^i{\bm{A}_b}{\bm{d}_{b,n}}\sum\limits_{j = 1}^{{L_{b,n}}} {\sqrt {\rho _{b,n}^j{p_{b,n}}} \bm{s}_{b,n}^j} }_{{\rm{ICI}}}
+ \underbrace {\bm{\upsilon} _{b,n}^i}_{{\rm{noise}}}  }
\end{aligned}
\end{equation}

As a result, the SINR of $i$th user at $n$th cluster in BS $b$ can be given by
\begin{equation}
\gamma _{b,n}^i =  \frac{{\rho _{b,n}^i{p_{b,n}}{{\left\| {\bm{h}_{b,n}^i{\bm{A}_b}{\bm{d}_{b,n}}} \right\|}^2}}}
{\left(\begin{array}{l} \sigma_n^2 + \sum\limits_{j = 1}^{{L_{b,n}}} {\rho _{b,n}^j} {p_{b,n}}{{\left\| {\bm{h}_{b,n}^i{\bm{A}_b}{\bm{d}_{b,n}}} \right\|}^2} \\ + \sum\limits_{j = 1,j \ne n}^N {{p_{b,j}}{{\left\| {\bm{h}_{b,n}^i{\bm{A}_b}{\bm{d}_{b,j}}} \right\|}^2}}  \end{array}\right)}
\end{equation}

Through the THz network capacity model given in \cite{THzw2016}, the achievable rate of $i$th user at $n$th cluster in BS $b$ can be given by
\begin{equation}
\bm{R}_{b,n}^i = \frac{W}{N}{\log _2}\left( {1 + \gamma _{b,n}^i} \right)
\end{equation}
where $W$ is the THz bandwidth employed at each SBS.

\subsection{THz Indoor Communication Channel Model}

In this subsection, THz band channel model is introduced specifically. In this paper, channel model of THz band is developed by using THz wave atmospheric transmission attenuation model and experiential water vapor continuum absorption.

In the existing work, many THz communication models have been proposed in \cite{THz2011,THz2014,THz2015,THz2016,THzpf2011,THzw2016}.
And all of those models include the LOS links and non-line-of-sight (NLOS) links.
NLOS links consist of reflected, scattered, and diffracted paths. However, scattered and diffracted paths are usually ignored for they only receive less power.
Because of the weak scattering ability, THz communication is sensitive to blockage of obstacles such as walls. In particular, blocking can lead to the difference between unblocked paths and blocked paths \cite{THzw2016}. In THz band, because the pathloss of NLOS link is much larger than that of LOS link, the influence of non-LOS link can be neglected when LOS link exists \cite{THz2016}. Therefore, the NLOS link of THz channel is very limited.

The wrok in \cite{THz2015} mainly studied the requirement of antenna number in THz-MIMO system.
And the work in \cite{THzpf2011} studied the influence on effectiveness of THz communication caused by distances from transmitters to receivers. For decreasing the computational complexity, we assume the distances are known by SBS.
So the channel gain of $i$th user on $n$th cluster of BS $b$ can be formulated as
\begin{equation}
\bm{h}_{b,n}^i\left( {f,\;d} \right) = \sqrt {{N_T}} (  \sqrt {\frac{1}{{\mathcal{PL}(f,d)}}} \Omega \bm{\alpha } \left( {\varphi _{b,n}^i} \right) )
\end{equation}
where $\mathcal{PL}(f,d)$ stands for the pathloss determined by THz frequency $f$ and distance $d$ between BS and user, $\Omega$ is the antenna gains,  $\bm{\alpha} \left( \varphi  \right)$ is the array steering vector.

In particular, path gain consists of spreading loss $\mathcal{L}_{sl}$ and molecular absorption loss $\mathcal{L}_{mal}$ which can not be neglected in THz band.
The spreading loss is caused by the expansion of electromagnetic wave as it propagates through various mediums.
The molecular absorption attenuation is a result of the collisions initiated by atmospheric gas or water molecules. More specific affect on atmospheric attenuation is studied in \cite{THz2011}. The authors researched the atmosphere molecular absorption by utilizing the HITRAN database.

Further, the pathloss of frequency $f$ suffers when traveling a distance $d$ can be expressed by:
\begin{equation}
{\mathcal{PL}}(f,d) = {\mathcal{L}_{sl}}(f,d) * \mathcal{L}_{mal}(f,d) = {\left( {\frac{{4\pi fd}}{c}} \right)^2}{e^{k(f)d}}
\end{equation}
or in dB:
\begin{equation}
\begin{aligned}
& { {\mathcal{PL}}(f,d)\left[ {dB} \right]  }
&& { = {\mathcal{L}_{sl}}(f,d)\left[ {dB} \right] + {\mathcal{L}_{mal}}(f,d)\left[ {dB} \right]  } \\
&&& { = 20{\log _{10}}\left( {\frac{{4\pi fd}}{c}} \right) + 10k(f)d{\log _{10}}e  }
\end{aligned}
\end{equation}
where $c$ is the speed of light in free space, $k(f)$ is frequency-dependent medium absorption coefficient

For uniform linear array, the array steering vector is only relevant to the array structure, which is given by
\begin{equation}
\bm{\alpha} \left( \varphi  \right) = \frac{1}{{\sqrt {{N_T}} }}{\left[ {1, \cdots ,{e^{j\pi \left[ {n\sin \varphi } \right]}}, \cdots ,{e^{j\pi \left[ {\left( {{N_T} - 1} \right)\sin \varphi } \right]}}} \right]^T}
\end{equation}
where $\varphi$ stands for angle of departure.

\subsection{Cache Model and Fronthaul Link }

In this subsection, cache model in THz band is introduced and the constraint on capacity of fronthaul link is given.
In this paper, we assume that the cache state information $c_{f,b}$  is predetermined.
Then, the binary variable of cache state information $c_{f,b}$ is given by
\begin{equation}
{c_{f,b}} = \left\{ {\begin{array}{*{20}{c}}
{\rm{1}}&{{\rm{if \ file }}f{\rm{ \ is \ cached \ by \ BS }}\ b}\\
{\rm{0}}&{{\rm{otherwise}}}
\end{array}} \right.
\end{equation}

If the file requested by the user has been cached by SBS, it can be retrieved directly from the cache rather than from MBS. Instead, uncached files need to be retrieved to MBS via the fronthaul link.
Further, the data rate of fronthaul link between MBS and SBS $b$ is is not difficult to obtain that
\begin{equation}
\bm{R}_b^{FH} = \sum\limits_{n = 1}^N {\sum\limits_{i = 1}^{{L_{b,n}}} {\bm{R}_{b,n}^i\left( {1 - F_{b,n}^i} \right)} }
\end{equation}

For simplicity, $F_{b,n}^i$ is cache efficiency coefficient brought by the long-term utility of the files needed by users from the cache, which can be written by
\begin{equation}
F_{b,n}^i = \frac{1}{{M_{b,n}^i}}\sum\limits_{f = 1}^{M_{b,n}^i} {{c_{f,b}}}
\end{equation}
where $M_{b,n}^i$ is the files number required by $i$th user on $n$th cluster of BS $b$.
The cache efficiency is an uncertain field, as the modeling of cache efficiency is still studied by many works. Nevertheless, this is beyond the scope of our paper since we focus on a caching strategy with a known cache efficiency.

Due to the high bandwidth of THz network, the cached files can not fully satisfy the requests. So some of the data needed by users need to be fetched from MBS through the fronthaul link and the capacity of the fronthaul link is often limited. Then, the fronthaul link capacity constraint can be formulate as:
\begin{equation}
\bm{R}_b^{FH} \le \bm{C}_b^{FH},\forall b \in \mathcal{B}
\end{equation}
It can be concluded that only when the fronthaul link capacity is infinite, can it reach the rate of cached network. The fronthaul link capacity restriction and no files needed by users in the local cache together limit the user's transmission rate.

\subsection{Power Consumption Model }

The total power consumption is composed of transmitting power and circuit power consumption, and the total power consumption of BS $b$ can be expressed as
\begin{equation}
{P_b} = P_b^c + \xi (\sum\limits_{n = 1}^N {{P_{b,n}}} )
\end{equation}
where $\xi$  is the inefficiency of the PA in THz networks, $P_b^c$ is the circuit power consumption of BS $b$ which is given by
\begin{equation}
P_b^c = {P_B} + {N_R}{P_R} + {N_T}{P_P} + {N_T}{P_A}
\end{equation}
where ${P_B}$ is the power consumption of baseband, ${P_R}$ is the power consumption of per RF chain, ${P_P}$ is the power consumption of per phase shifter, ${P_A}$ is the power consumption of per power amplifies.
For sub-connected structure, the number of phase shifters is ${N_T}$, not ${N_T}{N_R}$ in full-connected structure.

\subsection{Problem Formulation }

In this subsection, we propose an resource allocation problem by studied user clustering, hybrid precoding and power allocation subproblems.

EE is defined as the ratio of the profit brought by the long-term utility of capacity to the total power consumption.
Further, the system EE $\eta_{EE}$ can be written as
\begin{equation}
{\eta _{EE}} = \sum\limits_{b = 1}^B {\frac{{\sum\limits_{n = 1}^N {\sum\limits_{i = 1}^{{L_{b,n}}}\left( {1 + F_{b,n}^i} \right) {\bm{R}_{b,n}^i} } }}{{P_b^c + \xi (\sum\limits_{n = 1}^N {{P_{b,n}}} )}}}
\end{equation}

Based on the above system model, we define THz-NOMA network utility function to design user clustering, hybrid precoding and radio resource management in cached network with limited fronthaul link capacity.
Accordingly, the problem can be  modeled as follows
\begin{equation}
\max {\eta _{EE}}({\bm{A}_b},{\bm{d}_{b,n}},{p_{b,n}},\rho _{b,n}^i)  \\
\end{equation}
\begin{equation}
\begin{aligned}
&  {s.t.}
& &  C1: \sum\limits_{n = 1}^N {{{\left\| {{\bm{A}_b}{\bm{d}_{b,n}}} \right\|}^2}}  \le 1,\forall b \in B\\
&&& C2: \bm{R}_b^{FH} \le \bm{C}_b^{FH},\forall b \in B\\
&&& C3: \sum\limits_{n = 1}^N {{p_{b,n}}  \le {P_b^{\max }},\forall } b \in B\\
&&& C4: \sum\limits_{i = 1}^{{L_{b,n}}} {\rho _{b,n}^i  = 1,\forall } b \in B,n \in N
\end{aligned}
\end{equation}
where $C1$ denotes the normalization limitation of precoding vectors, $C2$ denotes fronthaul link capacity constraint,
$C3$ and $C4$ are the total power limitations. $C3$ guarantees that the sum of power on all beams does not exceed the maximum transmit power $P_b^{\max }$ of BS $b$. $C4$ guarantees that the sum of power control factors for all users on each beam is 1.

\section{User Clustering And Hybrid Precoding}

In this section, the user clustering and hybrid precoding are designed carefully. In particular, by effectively utilizing the transmission characteristics of NOMA, a fast convergent machine learning algorithm is proposed to realize user clustering.
Further, the hybrid precoding of THz-NOMA system is proposed to reduce the interference between users and achieve performance gains.

In THz communication, high antenna gains are advocated to remedy the high path loss. With the utilization of high-gain antennas, the transmission paths will become highly directional\cite{THz2014}.
The strong directional transmission characteristic of THz makes the user channel highly correlated, which is beneficial to the realization of NOMA, and makes the network obtain higher capacity and support more users.
Based on channel correlation, the user clustering and hybrid precoding are developed, they ensure that  the interference between beams is minimized and the network throughput is maximized.

\subsection{User Clustering}

In this subsection, an enhanced K-means based user clustering scheme is proposed.
In order to reduce the interference inter-beams, it is better to have strong correlation among users in a cluster. In large-scale MIMO systems, it is very difficult to find the optimal solution to the user clustering problem. In order to overcome the computational overhead of exhaustive search, some existing sub-optimal algorithms can solve the complex user clustering problem.

The authors of \cite{DLL2019} proposed a cluster head selected (CHS) scheme, in which cluster head are fixed, channel correlation is not fully considered, and is not the optimal solution.
In addition, match theory is introduced to solve complex user clustering problems \cite{b1,b2}, which has low complexity, but don't consider the learning characteristics of the algorithm itself.
The machine learning algorithm \cite{CJJ2018,KM-NOMA} which utilizes the correlation characteristics of CSI provides a new idea to study user clustering in MIMO-NOMA. The algorithm can converge to the global optimum approximately, but the initial cluster heads are set randomly without considering the network performance, the convergence speed is slow, and the user setting premise is not standardized.

In order to realize machine learning and clustering, it is very important to establish a quantitative feature set and a measure function about the target set.
Further, the channel correlation parameter of channel vectors ${\bm{h}_{{u_1}}}$ and ${\bm{h}_{{u_2}}}$ can be expressed as
\begin{equation}
M\left( {{\bm{h}_{{u_1}}},{\bm{h}_{{u_2}}}} \right) = \frac{{\left| {\bm{h}_{{u_1}}^H{\bm{h}_{{u_2}}}} \right|}}{{{{\left\| {\bm{h}_{{u_1}}^H} \right\|}_{\rm{2}}}{{\left\| {{\bm{h}_{{u_2}}}} \right\|}_{\rm{2}}}}}
\end{equation}

The K-means based algorithm is sensitive to initial clustering centers, which results in the fluctuation with the different initial clustering centers. To overcome this shortcoming, an enhanced K-means based user clustering scheme considering the initial cluster-head settings is proposed as described in Algorithm \ref{algorithm:1}, which can achieve faster convergence.

\begin{algorithm}
\caption{Enhanced K-means based User Clustering Scheme}
{\bf Input}   User set $U$, the number of cluster $N$, channel vectors $\bm{H}$; \\
{\bf 1$)$ Determine the initial cluster head set $\Theta^{\rm{'}}$}

\begin{algorithmic}[1]

\STATE  Random select a user as the first cluster head $\Theta_1^{\rm{'}}$
\FOR    {$n=2$ to $N$}
\STATE  ${\Theta _{n}^{\rm{'}}}{=}\mathop {\arg \max }\limits_{u \in U - \Theta } \sum\limits_{i = 1}^{n - 1} {M\left( {{\bm{h}_u},{\Theta _{i}^{\rm{'}}}} \right)} $;
\ENDFOR
\end{algorithmic}

{\bf 2$)$ The K-means iteration step (update cluster head)}

\begin{algorithmic}[1]
\REPEAT
\STATE  For each user $u$,
calculate $M\left( {{\bm{h}_u},{\Theta _{i}^{\rm{'}}}} \right),i=1, \cdots N $;
\STATE  The user $u$ belongs to the cluster with the smallest distance from it.
\STATE   Recalculating cluster centers
$\Theta _n^{\rm{'}} = \frac{1}{{\left| {{L_{b,n}}} \right|}}  \sum\limits_{u \in {U_{b,n}}}  {{\bm{h}_u}} $
\UNTIL the cluster members don't change;
\end{algorithmic}

\hspace*{0.02in} {\bf Output}   User set $U_{b,n}$ and channel vector of cluster head $\bm{h}_{b,n}$
\label{algorithm:1}
\end{algorithm}

The Algorithm \ref{algorithm:1} adopts the iterative updating method.
The advantage of Algorithm \ref{algorithm:1} is that users with less channel error should be selected as much as possible in the first step of selecting initial clustering centers, rather than random selection.
The $k$ initial clustering centers are selected according to the channel correlation parameter. The first cluster center ($n=1$) is selected by the random method. Then, the lower the channel correlation with the current $n$ cluster center users, the higher the probability that users are selected as the $n + 1$ cluster center, when $n$ initial cluster centers have been selected.

For complexity analysis, assuming that convergence is achieved through $t$ iterations, the asymptotic time complexity can be expressed as ${\rm O}(tNBU)$, but the complexity of cluster head selected scheme is ${\rm O}(tNBU^2)$ and the complexity of exhaustive search is ${\rm O}({BU^{2N + 1}})$.

\subsection{Analog Precoding}

In THz band network, precoding technology in MIMO can effectively compensate for path loss, but it requires lower complexity and larger antenna size. Therefore, efficient precoding in THz-MIMO is particularly important.
The sub-connection structure can greatly reduce the number of phase shifters corresponding to RF chain.

Using the classical two-stage scheme, we consider hybrid precoding, including baseband digital precoding and antenna subarray analog precoding.
For THz transceiver, the size and computation of the phase shifter are limited, and the power consumption of the phase shifter has an impact on the network performance. We consider $Q$-bits quantized phase shifters in each antenna subarray \cite{QPH}.

Based on cluster head obtained in last subsection, sub-connected analog precoding is designed to increase antenna gain according to the channel vectors of cluster heads.
consequently, the $s$th element of analog precoding vector of cluster $n$ of BS $b$ can be expressed by:
\begin{equation}
\bm{a}_{b,n}^{sub}(s) = \frac{1}{{\sqrt {N_T^{sub}} }}{e^{j\frac{{2\pi \omega }}{{{2^Q}}}}},u \in {\rm{\{ 1,2}}, \cdots ,N_T^{sub}{\rm{\}}}
\end{equation}
where quantized phase can be given by
\begin{equation}
\omega =\mathop {\arg \min }\limits_{\omega \in  \{ 0,1, \cdots ,{2^Q} - 1\} } \left| {\frac{{2\pi \omega }}{{{2^Q}}} - angle\left\{ {{h_{b,n}}(s)} \right\}} \right|
\end{equation}

\subsection{Digital Precoding}
After user clustering and analog precoding, digital Precoding is considered to eliminate MCI.
The baseband precoder $D_b$ changes the amplitude and phase of input complex symbols.
In baseband, we consider a low-dimensional channel equivalent matrix $\widehat {{\bm{H}_b}}$ based on cluster heads obtained by Algorithm \ref{algorithm:1}:
\begin{equation}
\widehat {{\bm{H}_b}} = \left[ {\bm{h}_{b,1}{\bm{A}_b},\bm{h}_{b,2}{\bm{A}_b}, \cdots ,\bm{h}_{b,N}{\bm{A}_b}} \right]
\end{equation}

Then, in order to solve the conventional MIMO-NOMA problem, a low-complexity zero-forcing precoding is proposed.
Without losing generality,  $\widehat {{\bm{H}_b}}$ is used for low-dimensional baseband precoding. The digital precoding matrix can be given by
\begin{equation}
\widehat {{\bm{D}_b}} = \left[ {{\bm{d}_{b,1}},{\bm{d}_{b,2}}, \cdots ,{\bm{d}_{b,N}}} \right] = {\widehat {{\bm{H}_b}}^H}{\left( {\widehat {{\bm{H}_b}}{{\widehat {{\bm{H}_b}}}^H}} \right)^{ - 1}}
\end{equation}

By introducing column power normalizing, the baseband precoding matrix can be expressed as
\begin{equation}
{\bm{D}_b} = \left[ {\frac{{{\bm{d}_{b,1}}}}{{{{\left\| {{\bm{A}_b}{\bm{d}_{b,1}}} \right\|}_2}}},\frac{{{\bm{d}_{b,2}}}}{{{{\left\| {{\bm{A}_b}{\bm{d}_{b,2}}} \right\|}_2}}}, \cdots ,\frac{{{\bm{d}_{b,N}}}}{{{{\left\| {{\bm{A}_b}{\bm{d}_{b,N}}} \right\|}_2}}}} \right]
\end{equation}

Up to now, user clustering and hybrid precoding have been designed.
In order to decode smoothly, users need to be indexed and sorted after user clustering and hybrid precoding.
For each cluster, reorder user according to equivalent gain as the following rule:

\begin{equation}
{\left\| {\bm{\overline {h}}_{b,n}^1{\bm{A}_b}{\bm{d}_{b,n}}} \right\|_2} \ge {\left\| {\bm{\overline {h}} _{b,n}^2{\bm{A}_b}{\bm{d}_{b,n}}} \right\|_2} \ge  \cdots  \ge {\left\| {\bm{\overline {h} } _{b,n}^{{L_{b,n}}}{\bm{A}_b}{\bm{d}_{b,n}}} \right\|_2}
\end{equation}

So far, hybrid precoding has been carefully designed to eliminate interference. In addition, how to utilize limited energy has not been solved under the condition of low transmission power and huge data rate of THz network. In the next section, power optimization for THz-NOMA network will be studied to max the EE in (20).

\newcounter{TempEqCnt}
\setcounter{TempEqCnt}{\value{equation}}
\setcounter{equation}{29}
\begin{figure*}[hb]
\begin{equation}
\bm{\widehat y} _{b,n}^i =
 \underbrace {\bm{\bar h}_{b,n}^i{\bm{A}_b}{\bm{d}_{b,n}}\sqrt {{p_{b,n,i}}} \bm{s}_{b,n}^i}_{{\rm{desired}}\;{\rm{signal}}} + \underbrace {\bm{\bar h}_{b,n}^i{\bm{A}_b}\sum\limits_{j = 1,j \ne n}^N {{\bm{d}_{b,j}}} {\bm{x}_{b,j}}}_{{\rm{MCI}}}
 + \underbrace {\bm{\bar h}_{b,n}^i{\bm{A}_b}{\bm{d}_{b,n}}\left( {\sum\limits_{j = 1}^{i - 1} {\sqrt {{p_{b,n,j}}} \bm{s}_{b,n}^j}  + \phi \sum\limits_{j = i + 1}^{{L_{b,n}}} {\sqrt {{p_{b,n,j}}} \bm{s}_{b,n}^j} } \right)}_{{\rm{residual}}\;{\rm{ICI}}} + \underbrace {\bm{\upsilon} _{b,n}^i}_{{\rm{noise}}}
\label{3030}
\end{equation}
\end{figure*}

\section{Power Optimization}

Terahertz radiation power is low, so it is a challenge to meet the carrier power requirements of communication.
Therefore, it is necessary to study the THz network combined with power optimization of multi-antenna technology to alleviate the problems of low output power and low energy conversion efficiency of THz source.
Nowadays, there are many studies on power optimization in MIMO-NOMA network such as \cite{THzbpb2019}\cite{EE2017}\cite{MIMO-NOMA}.
The introduction of cache guarantees network capacity.
However, the capacity limitation of the fronthaul links also poses a challenge to power optimization in THz MIMO-NOMA network, since the power of different users is coupled.
To solve so intractable problem, a distributed resource allocation via ADMM is proposed considering imperfect SIC.

After the user clustering and hybrid precoding, power allocation is utilized to enhance the EE.
Given the reordered user gains $\bm{\overline h} _{b,n}^i$, hybrid precoding matrixs ${\bm{A}_b}$ and ${\bm{D}_{b}}$, the power allocation subproblem can be expressed as
\setcounter{equation}{28}
\begin{equation}
\begin{array}{c}
\max {\eta _{EE}}({p_{b,n}},\rho _{b,n}^i)\\
s.t.\;{\rm{ }}C2,C3,C4
\end{array}
\label{gs:wt}
\end{equation}

For power domain NOMA network, SIC technique is introduced to reduce intra-cluster interference.
In this way, users with strong channel gain can remove the interference caused by users with weak channel gain.
The SIC technique and effective user clustering scheme together effectively reduce the multiple interference for MIMO-NOMA network.
However, in the practical scenario, it may not be practical to assume the perfect SIC at the user terminal. Because there are still some serious implementation problems to SIC, such as limited computing power and error propagation \cite{iSIC}.
After decoding with imperfect SIC, the received signal of $i$th user at $n$th cluster in BS $b$ is represented as (\ref{3030}) at the bottom of this page, where $\phi $ is the cancellation error arising from imperfect SIC, ${p_{b,n,i}} = \rho _{b,n}^i{p_{b,n}}$ is the transmit power of $i$th user at $n$th cluster in BS $b$ expressed in terms of the power control factor and  transmit power of the cluster.

\setcounter{equation}{30}

Then, the SINR of $i$th user at $n$th cluster in BS $b$ can be rewritten by
\begin{equation}
\widehat \gamma _{b,n}^i{\rm{ = }}\frac{{{p_{b,n,i}}{{\left\| {\bm{\bar h}_{b,n}^i{\bm{A}_b}{\bm{d}_{b,n}}} \right\|}^2}}}
{\begin{array}{l}
 \left( {\sum\limits_{j = 1}^{i - 1} {{p_{b,n,j}}}  + \phi \sum\limits_{j = i + 1}^{{L_{b,n}}} {{p_{b,n,j}}} } \right){{\left\| {\bm{\bar h}_{b,n}^i{\bm{A}_b}{\bm{d}_{b,n}}} \right\|}^2} \\
 +  \sum\limits_{m = 1,m \ne n}^N {{p_{b,m}}{{\left\| {\bm{\bar h}_{b,n}^i{\bm{A}_b}{\bm{d}_{b,m}}} \right\|}^2}} \sigma _n^2 + \end{array}} .
\end{equation}

Furthermore, the EE of THz system can be rewritten by
\begin{equation}
\begin{aligned}
&    {\widehat \eta  _{EE}}
&&   = \sum\limits_{b = 1}^B {{{\sum\limits_{n = 1}^N {\sum\limits_{i = 1}^{{L_{b,n}}} \frac{1}{P_b} {\left( {1 + F_{b,n}^i} \right)}   \bm{\widehat R} _{b,n}^i  } }}  } \\
&&&  = \sum\limits_{b = 1}^B {\sum\limits_{n = 1}^N {\sum\limits_{i = 1}^{{L_{b,n}}} {\frac{{\frac{W}{N}\left( {1 + F_{b,n}^i} \right){{\log }_2}\left( {1 + \hat \gamma _{b,n}^i} \right)}}{{P_b^c + \xi (\sum\limits_{n = 1}^N {{P_{b,n}}} )}}} } }
\end{aligned}
\label{gs:ee}
\end{equation}

Specifically, we consider a dinkelbach-style algorithm to transform the original nonlinear optimization problem of fractional program form into an equivalent parameterized subtractive form function:
The utility function of problem (\ref{gs:wt}) can be rewritten with a subtractive-form by
\begin{equation}
f\left( {\widehat \eta _{EE}  ^ *} \right) = \sum\limits_{b = 1}^B {\sum\limits_{n = 1}^N {\sum\limits_{i = 1}^{{L_{b,n}}} {\left( {1 + F_{b,n}^i} \right)\bm{\widehat R} _{b,n}^i} }  - {\widehat \eta _{EE} ^ *} \sum\limits_{b = 1}^B { {P_b}    } }
\end{equation}
where parameter $\widehat \eta  _{EE}^ *$ is an auxiliary variable and is used to scale the weight of ${P_b}$. Clearly,
$f( {\widehat \eta _{EE}  ^ *})$ is a convex function with respect to $ \widehat \eta  _{EE} ^ * $. Therefore, solving (\ref{gs:wt}) is equivalent to finding the roots of equation $f( {\widehat \eta _{EE}  ^ *}) = 0$.
Then, the resource allocation problem of (\ref{gs:wt}) can be rewritten by
\begin{equation}
\mathop {\max }\limits_{{p_{b,n,i}}} \sum\limits_{b = 1}^B {\sum\limits_{n = 1}^N {\sum\limits_{i = 1}^{{L_{b,n}}} {\left( {1 + F_{b,n}^i} \right) \bm{\widehat R} _{b,n}^i} }  - \widehat \eta  _{EE}^ * \sum\limits_{b = 1}^B { \sum\limits_{n = 1}^N {\xi {P_{b,n}}}} }
\label{gs:wt2}
\end{equation}
\begin{equation}
\begin{aligned}
& {s.t.}
&&  C2': \bm{\widehat R}_b^{TH} \le \bm{C}_b^{TH},\forall b \in B\\
&&& C5: {\sum\limits_{n = 1}^N {\sum\limits_{i = 1}^{{L_{b,n}}} {{p_{b,n,i}} \le P_b^{\max }} } ,\forall b \in B}
\end{aligned}
\end{equation}

Furthermore, distributed resource allocation via ADMM is utilized to work out optimal power assignment problem.
More specific introduction about the ADMM method can be captured in \cite{ADMM-JSAC2016}\cite{ADMM-JSAC2019}. Before utilizing the unscaled-form ADMM to design the power assignment problem, we firstly present two parameters, i.e, $X$ and $Z$. They are both the auxiliary vectors. $X$ denotes the all distributed power elements of each user. $Z$ is a global auxiliary vector and its each factor correspond to one in $X$.
Moreover, we define $\Gamma$ as the set of variable vectors, which satisfy constraint $C2'$.
Then, the indicator function is introduced by $g\left( Z\right) = 0$ when $Z \in \Gamma$, otherwise $g\left( Z \right) =  + \infty $.
On the basis of the above introduction, the power optimization problem (\ref{gs:wt2}) is changed to
\begin{equation}
\begin{array}{c}
\mathop {\min }\limits_{X} \left[  \begin{array}{l}
\widehat \eta  _{EE}^ * \sum\limits_{b = 1}^B { \sum\limits_{n = 1}^N {\xi {P_{b,n}}}} + g\left( Z \right) \\
- \sum\limits_{b = 1}^B \sum\limits_{n = 1}^N {\sum\limits_{i = 1}^{{L_{b,n}}} {\left( {1 + F_{b,n}^i} \right) \bm{\widehat R} _{b,n}^i} }
  \end{array}  \right] \\
s.t. {\rm{ }} X - Z = 0
\end{array}
\end{equation}

In the unscaled form, the augmented Lagrangian $L_\mu $ is given as follows
\begin{equation}
\begin{aligned}
&   {L_\mu  } =
&&   \widehat \eta  _{EE}^ * \sum\limits_{b = 1}^B { \sum\limits_{n = 1}^N {\xi {P_{b,n}}}}
- \sum\limits_{b = 1}^B \sum\limits_{n = 1}^N {\sum\limits_{i = 1}^{{L_{b,n}}} {\left( {1 + F_{b,n}^i} \right)\bm{\widehat R} _{b,n}^i} } \\
&&& + g\left( Z \right) + {\lambda ^T}\left( {X - Z} \right) + \frac{\mu }{2}\left\| {X - Z} \right\|_2^2
\end{aligned}
\end{equation}
where $\lambda$ is the vector of dual variable, $\mu  > 0 $ is the predefined augmented Lagrangian parameter. In order to
work out the power assignment problem, we design the following steps
\begin{equation}
\begin{array}{l}
{X^{i + 1}} = \mathop {\arg \min }\limits_X \left[ \begin{array}{l}
\hat \eta _{EE}^*\sum\limits_{b = 1}^B {\sum\limits_{n = 1}^N {\xi {P_{b,n}}} } \\ - \sum\limits_{b = 1}^B {\sum\limits_{n = 1}^N {\sum\limits_{i = 1}^{{L_{b,n}}} {\left( {1 + F_{b,n}^i} \right)\bm{\widehat R}_{b,n}^i} } } \\
 + {({\lambda ^i})^T}\left( {X - {Z^i}} \right) + \frac{\mu }{2}\left\| {X - {Z^i}} \right\|_2^2
\end{array} \right] \\
{Z^{i + 1}} = \mathop {\arg \min }\limits_Z   \left[
{(\lambda^i) ^T}\left( {X^{i+1} - Z} \right) + \frac{\mu }{2}\left\| {X^{i + 1} - Z} \right\|_2^2 \right] \\
{\lambda ^{i + 1}} = {\lambda ^i} + \mu  \left( {{X^{i + 1}} - {Z^{i + 1}}} \right)
\end{array}
\label{gs:xy}
\end{equation}
where $i$ is the iteration index.
Then, the ADMM power allocation scheme for THz-NOMA-MIMO downlink network is described in detail in Algorithm \ref{algorithm:2} below.
\begin{algorithm}[!ht]
\vspace{3mm}
\caption{Distributed Energy-efficient Algorithm for Power Allocation via ADMM}
\begin{algorithmic}[1]
\STATE  \textbf{Initialize} Set iteration index $t=1$, auxiliary variable $\eta _{EE} ^*=0$, stop standard $\theta ^ * $ and the power is allocated equally
\REPEAT
\STATE  Set $i=1$
\REPEAT
\STATE update $X^{i + 1}$ according to (\ref{gs:xy}) ;
\STATE update $Z^{i + 1}$ according to (\ref{gs:xy}) ;
\STATE update $\lambda ^{i + 1}$ according to (\ref{gs:xy}) ;
\STATE $i=i+1$
\UNTIL converge
\STATE calculate $\eta _{EE} ^* = \eta _{EE} ^{t+1}$ according to (\ref{gs:ee});
\STATE calculate  $\theta=\eta _{EE} ^{t+1}-\eta _{EE} ^t$
\STATE $t=t+1$
\UNTIL $\theta  \le {\theta ^ * }$
\end{algorithmic}
\label{algorithm:2}
\end{algorithm}

For complexity analysis, assuming that convergence is achieved through $T$ iterations.
In Algorithm 2, we need to update the power for each iteration. The calculation of (32) for $B$ BSs with $U$ users needs $BU$ operations. Suppose $X$ converges within ${T^*}$ iterations.
The updates of $X$ needs $O({T^*}BU)$ operations.
Therefore, the total complexity of Algorithm 2 is $O(T((1 + {T^*})BU))$.

\section{Numerical Simulation And Analysis}
In this section, we provide the simulation results to verify the performance of user clustering, hybrid precoding and power allocation methods.
Specifically, we compare the proposed user clustering Algorithm \ref{algorithm:1} with the machine learning algorithm \cite{CJJ2018} and cluster head selection algorithm \cite{DLL2019}. In particular, the effects of different hybrid precoding schemes on total EE and sum rate are compared. Furthermore, the comparison of EE with different cache efficiency coefficients and forward link capacity constraints is also shown.
In addition, the advantages of NOMA system are verified by comparing OMA system.

In this work, it is important to select frequency intelligently to avoid the spectrum with path loss peak. In order to provide large channel capacity with low path loss, we choose 0.34 THz carrier frequency in directional propagation to avoid path loss peak \cite{THz2016}.
In downlink THz-NOMA system, we assume that MBS are in the center, SBS is in the MBS coverage, and all users are randomly allocated within the coverage of their associated BS.
In this simulation, the radius of SBS is set to 5 meters, the minimum distance between user is 0.1 meter, and the minimum distance between BS and user is 0.5 meters.
The system bandwidth $W$ is set to 10 GHz.
The AWGN power spectral density $N_0$ is set to -174 dBm/Hz.
To reduce the computational complexity, we set the number of SBS and users per BS to $B=2$; $U_b=15$, respectively.
The cache efficiency used in the simulation is 0.3.
The maximum transmit power of SBS is defined as 5W.
The power consumption of baseband $P_{B}$ is 200mW,
the power consumption of per RF chain $P_{R}$ is 160mW ;
the power consumption of 4-bit phase shifter $P_{P}$ is 40mW;
and the power consumption of the PA $P_{P}$ is 20 mW.
The power consumption of the phase shifter with different quantization bits is different, which consumes 10mW per bit.
Moreover, the inefficiency of the PA $\xi$ is $1/0.38$.
In the process of atmospheric attenuation, the contribution of water vapor molecules to the performance of THz channels is significantly greater than that of other air molecules.  For simplification, we only consider the effect of water molecules in the horizontal term of atmospheric absorption, which will be used in our simulation. The parameters of frequency absorption coefficient $K(f)$ are shown in \cite{THz2011}.

\begin{figure}[t]
        \centering
        \includegraphics*[width=8cm]{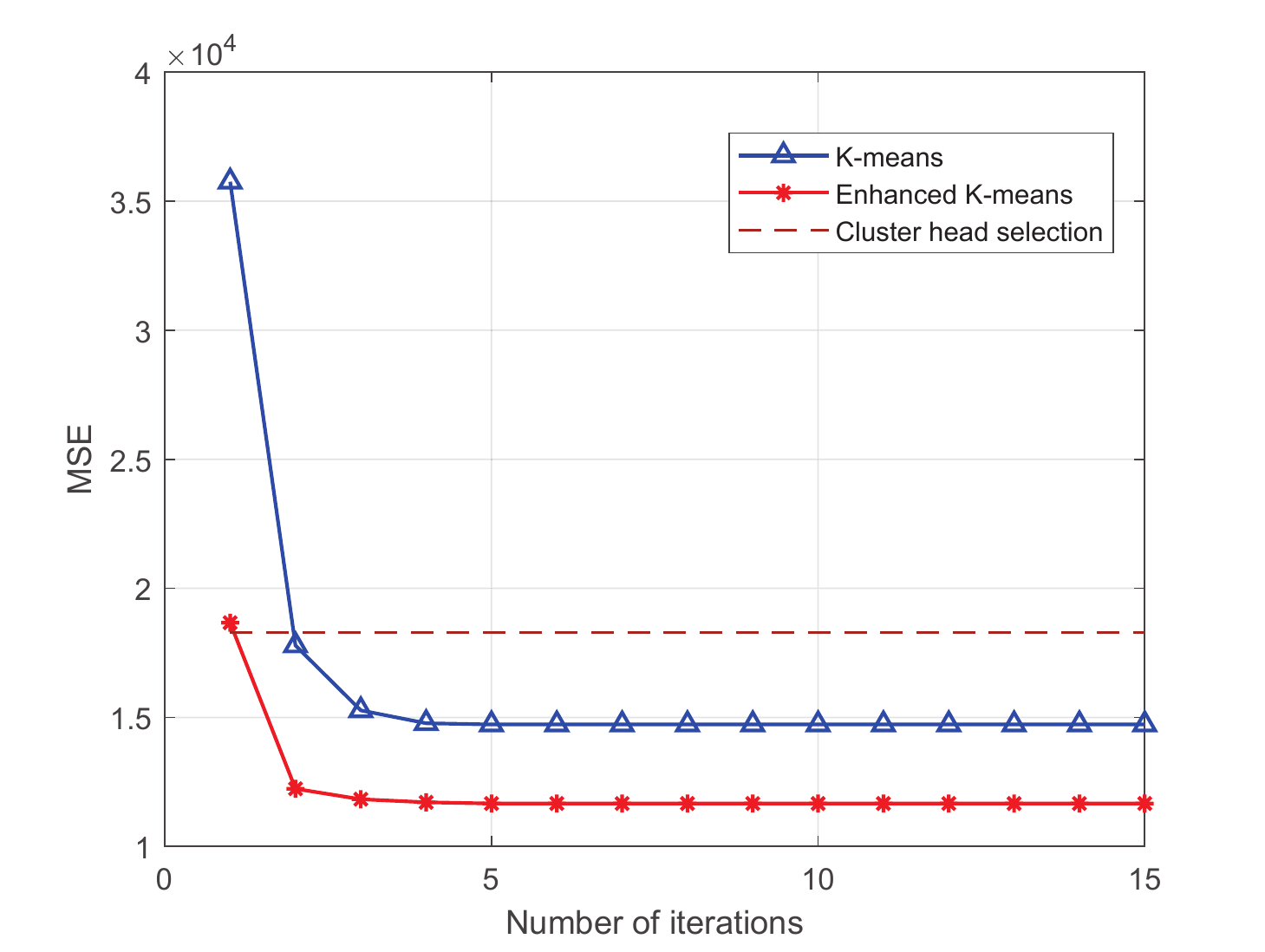}
       \caption{MSE comparison versus number of iterations under different user cluster algorithm.}
        \label{fig:333}
\end{figure}
Fig. \ref{fig:333} studies the comparison of mean square error (MSE) with the number of iterations under different user clustering algorithms.
The goal of user clustering is to minimize the sum of MSE among all clusters, that is to maximize the channel correlation of users in each cluster.
The convergence of MSE under K-means and proposed enhanced K-means for user clustering can be seen in the figure.
Obviously, the proposed Algorithm \ref{algorithm:1} can achieve convergence faster than K-means algorithm.
As shown in the Fig. \ref{fig:333}, the MSE of the proposed Algorithm \ref{algorithm:1} at the first iteration is close to the cluster head selection algorithm, much lower than the MSE of K-means algorithm.
Because the cluster head selection algorithm does not iterate after the cluster head is determined, the MSE remains at a fixed value.

\begin{figure}[t]
        \centering
        \includegraphics*[width=8cm]{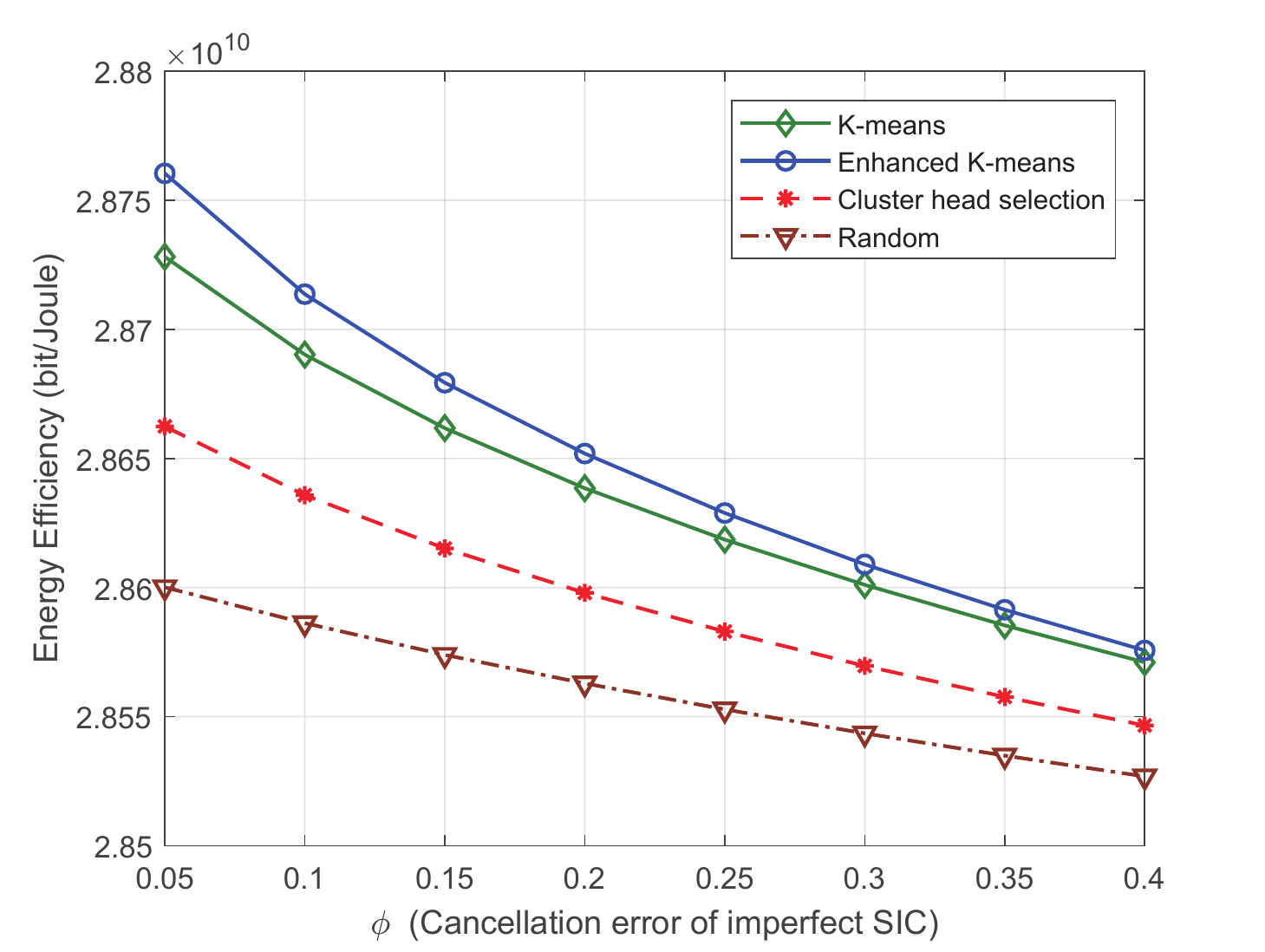}
       \caption{The total EE comparison versus cancellation error arising from SIC under different user cluster algorithm.}
        \label{fig:555}
\end{figure}
Fig. \ref{fig:555} studies EE of user clustering scheme based on different algorithms.
Specifically, considering cancellation error $\phi$ of imperfect SIC, with the increase of cancellation error, the intra-cluster interference received by users increases significantly, resulting in the decrease of EE of the system.
Fig. \ref{fig:555} shows that the proposed algorithm can achieve higher EE than other algorithms, which shows that the proposed algorithm can effectively divide users.
For K-means user clustering, although user centers can be found well according to user's CSI, the initial random selection of cluster centers will affect the convergence of the system and then affect the EE of the system.
For random clustering, the EE of THz-NOMA-MIMO system decreases seriously. This is because of serious inter-cluster interference, it is difficult for random clustering to get more benefits from NOMA scheme.
Therefore, in the later simulation analysis, the enhanced K-means algorithm is used for user clustering and the value of $\phi$ is set to 0.005.

\begin{figure}[t]
        \centering
        \includegraphics*[width=8cm]{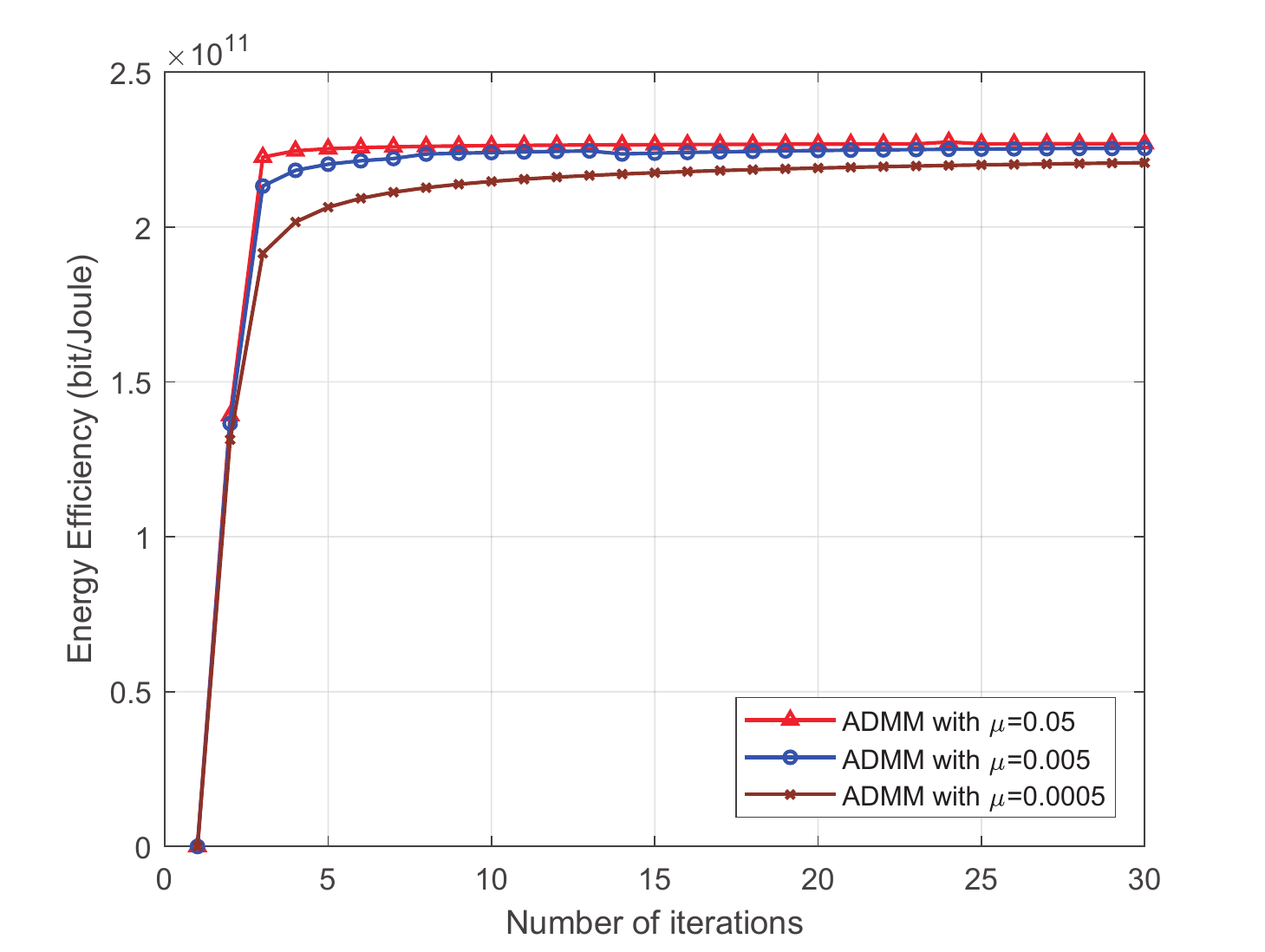}
       \caption{Convergence in terms of the total EE via ADMM of different $\mu$.}
        \label{fig:444}
\end{figure}
Fig. \ref{fig:444} shows the EE  comparison against different $\mu$ implemented by ADMM algorithm for power optimization.
In the simulation, the number of transmitting antennas is 64, and the hybrid precoding adopts partial connection scheme.
As shown in the Fig. \ref{fig:444}, the EE increases with the number of iterations until convergence.
The EE of the system eventually converges from $2.2\times {10^{11}}$ bps/J/Hz to $2.3\times{10^{11}}$ bps/J/Hz.
The system EE tends to be stable after 10 iterations, which verifies the convergence of the Algorithm \ref{algorithm:2} for power optimization problem discussed in Section IV.
It is noteworthy that EE converges faster as $\mu$ increases.
This verifies the effect of $\mu$ on the convergence of the ADMM algorithm for power allocation.
In the later simulation, the value of $\mu$ is set to 0.05 in order to converge faster.

\begin{figure}[t]
        \centering
        \includegraphics*[width=8cm]{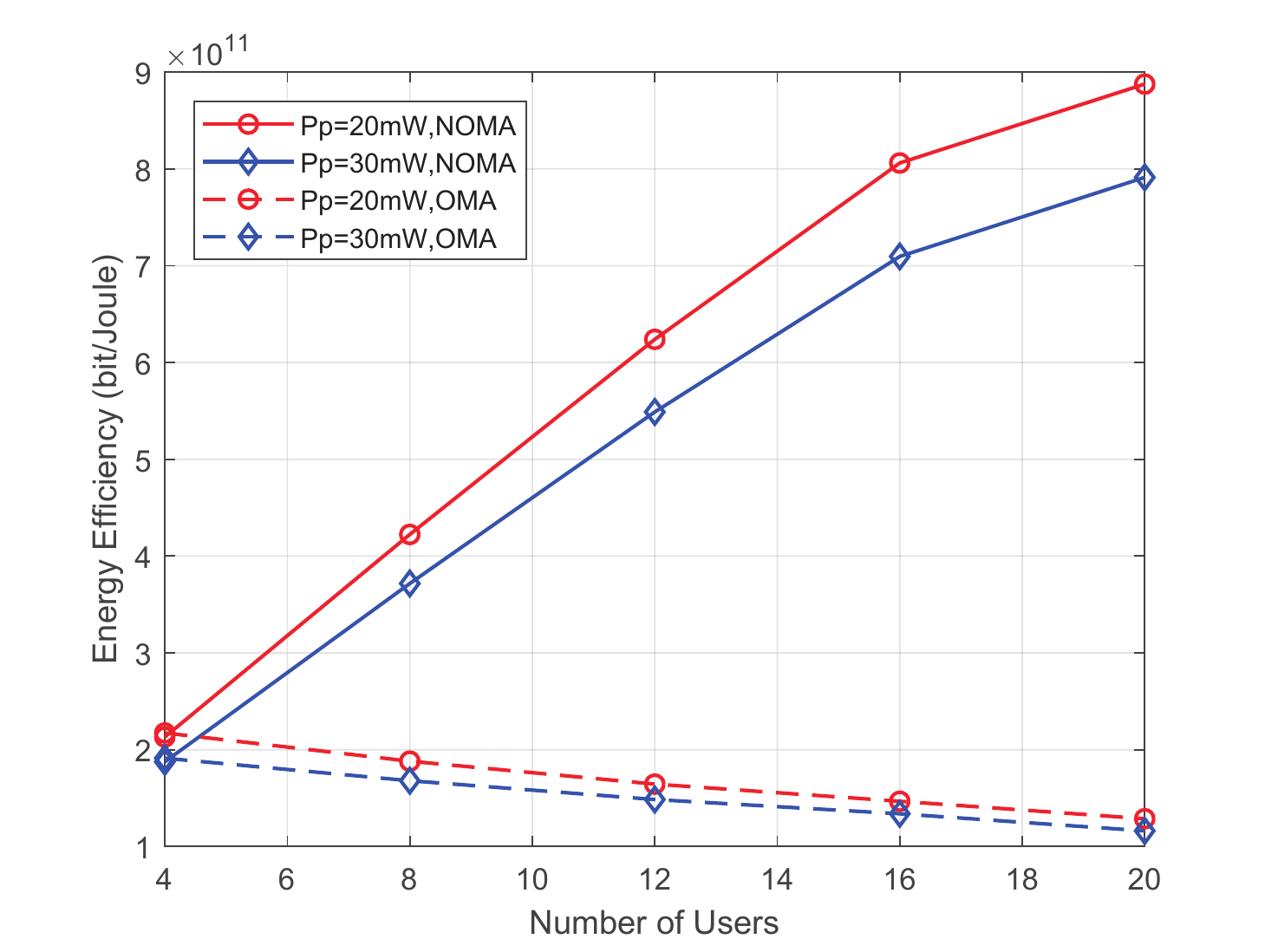}
       \caption{The total EE comparison versus number of users under different power of per phase shifter and different  multiple access method  of NOMA or OMA.}
        \label{fig:666}
\end{figure}
Fig. \ref{fig:666} shows the EE  comparison against different number of users of considered 4 schemes, where the number of clusters is 4, the number of users increased from 4 to 20 and the hybrid precoding adopts sub-connection scheme.
We represent OMA with FDMA, in which each channel can only be assigned to one user at a time.
It can be seen that NOMA system improves the EE of the system significantly compared with OMA system.
The system EE of NOMA-MIMO system increases significantly with the increase of users, while OMA-MIMO system is the opposite.
This is because the overlay of multiple users in the same cluster increases the load of the network, so that more users can get more bandwidth, which greatly increases the system capacity.
In addition, the comparison of system EE is also taken into account under different power consumption $P_p$ of phase shifter.
It can be seen that for NOMA-MIMO system and OMA-MIMO system, the EE realized when $P_p$ is 20 mW is larger than that when $P_p$ is 30 mW.
This is because the power consumption caused by the phase shifter will directly affect the system EE.

\begin{figure}[t]
        \centering
        \includegraphics*[width=8cm]{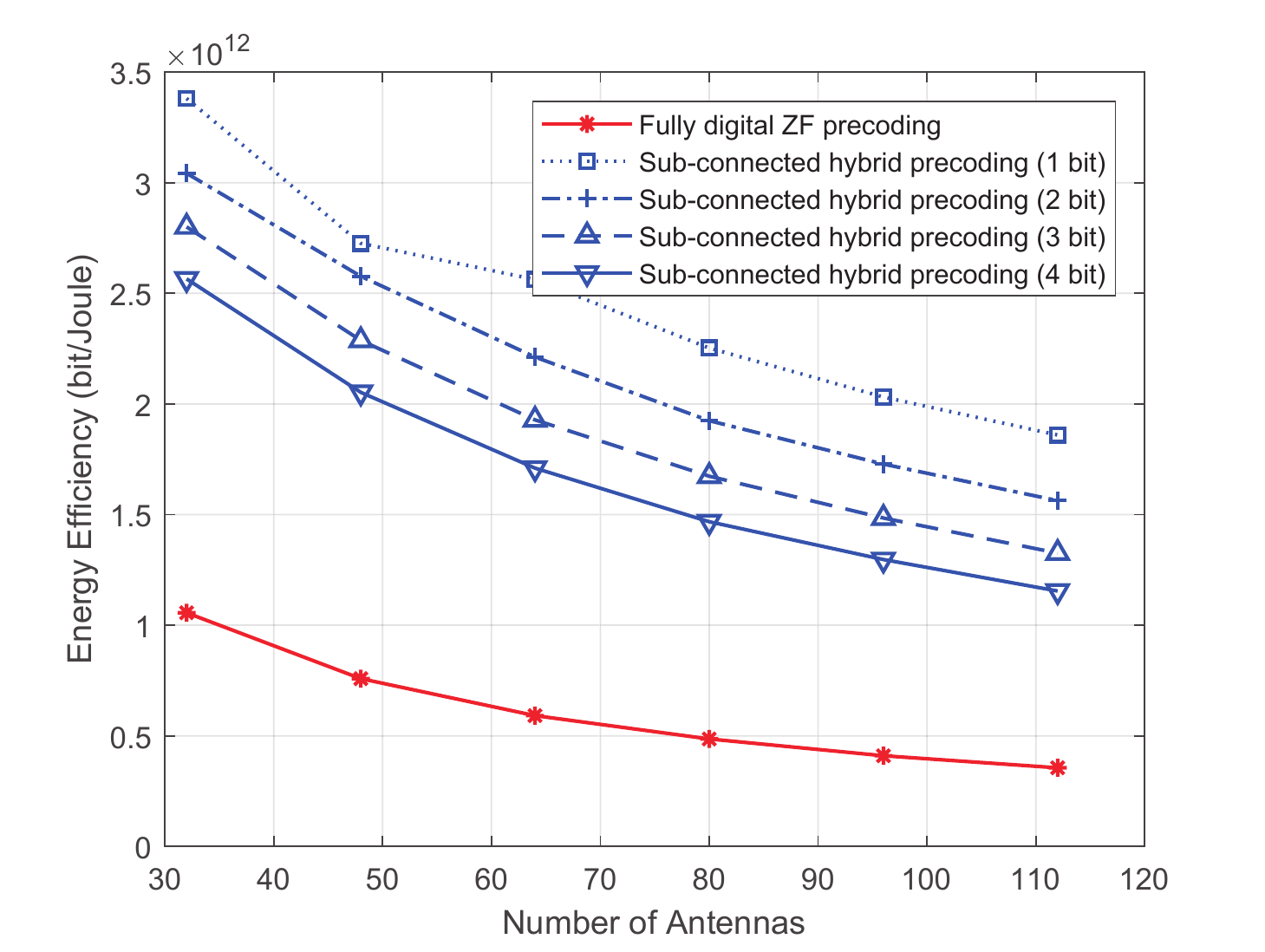}
      \caption{The total EE comparison versus number of antennas under different precoding schemes.}
        \label{fig:777}
\end{figure}

Fig. \ref{fig:777} shows EE against the different number of antennas under different precoding schemes, where the number of clusters is 2, the number of users is 15.
It can be seen that the hybrid precoding of NOMA-MIMO system with sub-connected structure can achieve higher EE than digital ZF precoding.
Moreover, the impact of quantization bits of phase shifter is also validated.
As can be seen from Fig. \ref{fig:777}, the THz-NOMA system with low bits of phase shifter can achieve higher EE than that with high bits of phase shifter.
This is because the higher quantization bits leads to higher power consumption and affects energy efficiency.

\begin{figure}[t]
        \centering
        \includegraphics*[width=8cm]{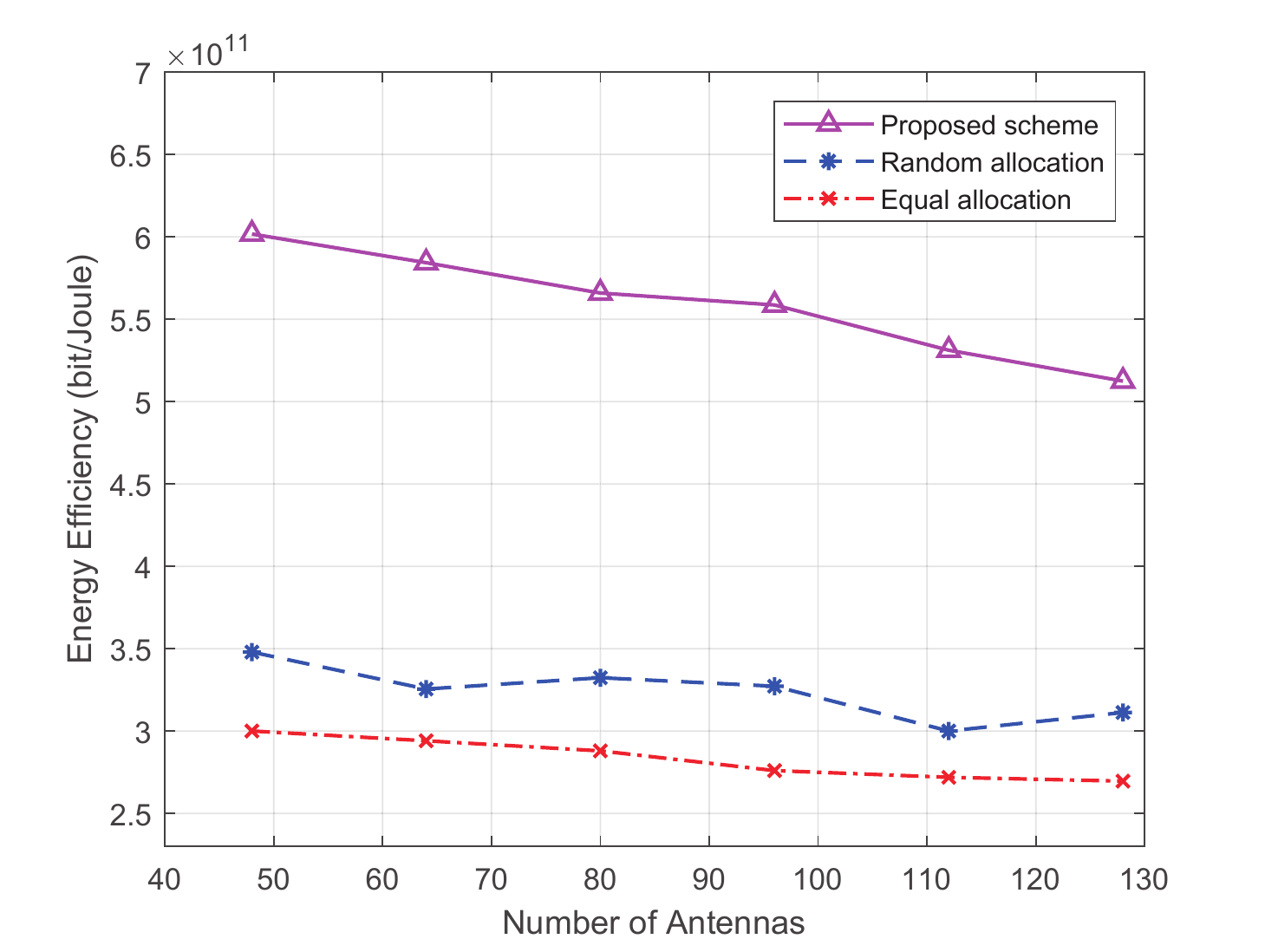}
     \caption{ The total EE comparison versus number of antennas under different power allocation algorithms.}
        \label{fig:888}
\end{figure}

Fig. \ref{fig:888} depicts the EE performance under different power allocation algorithms, where the number of clusters is 2, the number of users is 15.
As the number of antennas increases, the EE of THz-NOMA systems with sub-connected hybrid precoding decreases.
From the Fig. \ref{fig:888}, it can be seen that our proposed power allocation algorithm can achieve higher EE than random power allocation and equal power allocation.

\begin{figure}[t]
        \centering
        \includegraphics*[width=8cm]{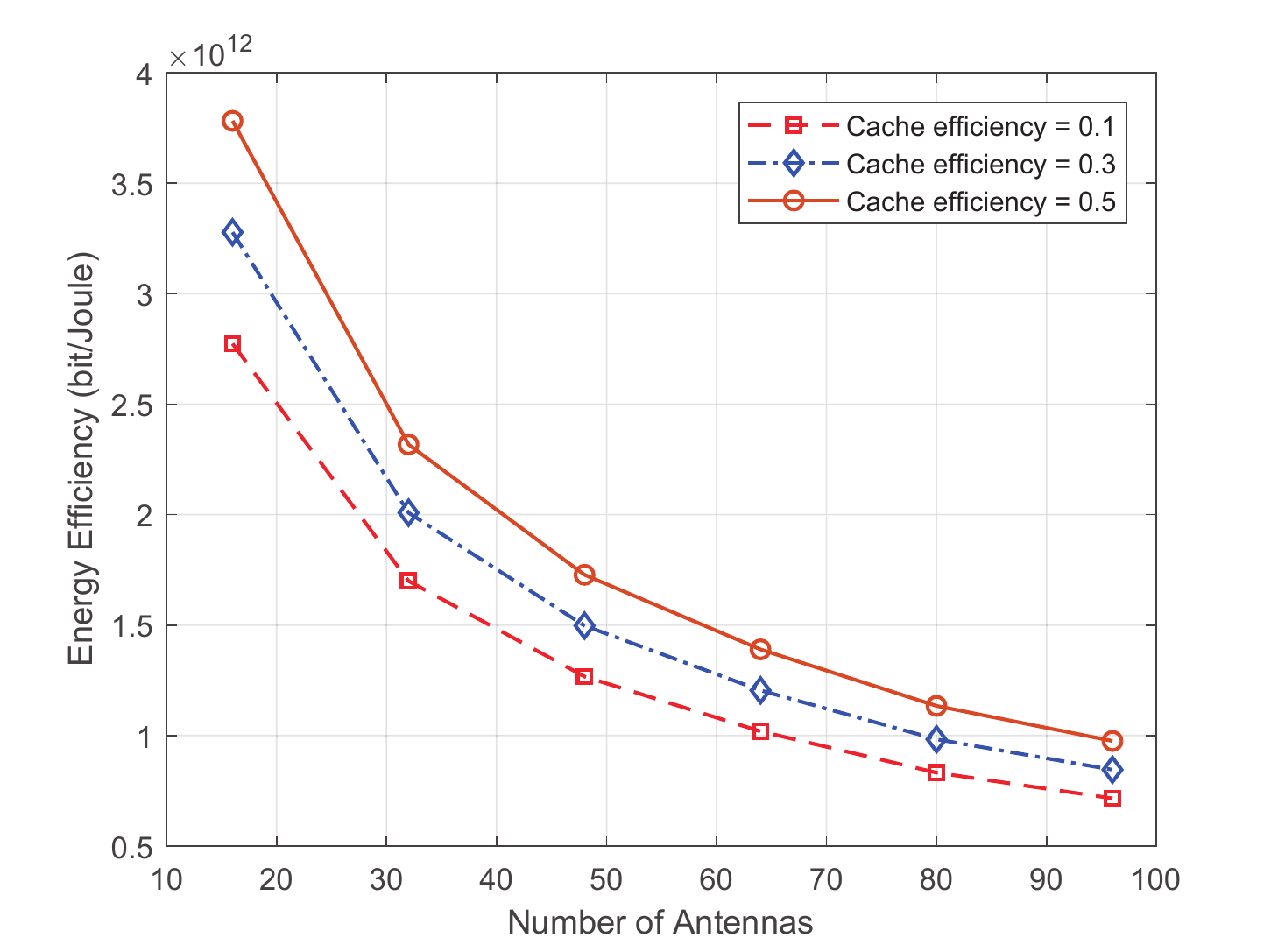}
       \caption{The total EE comparison versus number of antennas under different cache efficiency.}
        \label{fig:1010}
\end{figure}

Fig. \ref{fig:1010} shows the impact of different cache efficiency on the EE of the system.
When the cache efficiency increases gradually, the value of EE increases with the increase of $F$. This is because the introduction of local cache files greatly increases the user's sum rate.
Users can directly obtain the required files through the local cache at BS.  When more files are retrieved from the cache, the profit brought by the long-term utility of capacity becomes more and more obvious.
In addition, the EE decreases with the increase of the number of transmitting antennas, due to the power consumption of transmitting antennas, which has been discussed in Fig. \ref{fig:777} and Fig. \ref{fig:888}.

\section{Conclusions}

In this paper, we study the maximization of EE problem in THz-NOMA-MIMO systems and design user clustering, hybrid precoding and power optimization strategies.
A fast convergence scheme for user clustering in NOMA-MIMO system by using enhanced K-means machine learning algorithm is proposed.
Considering the power consumption and implementation complexity, the hybrid precoding scheme based on the sub-connection structure is  adopted.
Considering the fronthaul link capacity constraint, we design a distributed ADMM algorithm for power allocation to maximize the EE of THz-NOMA system with imperfect SIC.
The simulation results show that the proposed user clustering scheme can achieve faster convergence and higher EE, the design of the hybrid precoding of the sub-connection structure can achieve lower power consumption and power optimization can achieve a higher EE for the THz cache-enabled network.

\end{document}